\documentclass[]{jfm}
\usepackage{graphicx}
\usepackage{siunitx}
\usepackage{amsmath}
\usepackage{epstopdf, epsfig}
\usepackage{upgreek}
\usepackage[caption=false]{subfig}
\usepackage{array}
\usepackage{microtype}

\usepackage[usenames,dvipsnames]{xcolor}
\usepackage{booktabs}
\usepackage{textcomp} 
\usepackage{scrhack} 
\usepackage{xspace} 
\usepackage{mparhack} 
\usepackage{tabularx} 

\shorttitle{Marangoni spreading of binary drops} 
\shortauthor{B. F. van Capelleveen, R. B. J. Koldeweij, D. Lohse and C. W. Visser}

\title{Marangoni-driven spreading of miscible liquids in the binary drop geometry}

\author{Robin B. J. Koldeweij\aff{1,}\aff{2} \unskip\footnotemark[1],
Bram F. van Capelleveen\aff{1} \aunote{R. B. J. Koldeweij and B. F. van Capelleveen contributed equally to this work}, 
  Detlef Lohse\aff{1,}\aff{3}
  \and Claas Willem Visser\aff{1,}\aff{4} \corresp{\email{c.visser@utwente.nl}} 
}

\affiliation{\aff{1}Physics of Fluids Group \& Max Planck Center Twente for Complex Fluid Dynamics, Department of Science and Technology, J. M. Burgers Center for Fluid Dynamics, University of Twente, 7500 AE Enschede, The Netherlands
\aff{2} Equipment for Additive Manufacturing, TNO, 5612 AP Eindhoven, The Netherlands
\aff{3} Max Planck Institute for Dynamics and Self-Organization, 37077 G\"{o}ttingen, Germany
\aff{4} Fluid Mechanics for Functional Materials group, Department of Thermal and Fluid Engineering, Faculty of Engineering Technology, University of Twente, 7500 AE Enschede, The Netherlands
}

\begin{document}

\maketitle

\begin{abstract}
When two liquids with different surface tensions come into contact, the liquid with lower surface tension spreads over the other. This Marangoni-driven spreading has been studied for various geometries and surfactants, but the dynamics of the binary geometry (drop-drop) has hardly been quantitatively investigated, despite its relevance for drop encapsulation applications. Here we use laser-induced fluorescence (LIF) to temporally resolve the distance $L(t)$ over which a low-surface-tension drop spreads over a miscible high-surface-tension drop. $L(t)$ is measured for various surface tension differences between the liquids and for various viscosities, revealing a power-law $L(t)\sim t^{\alpha}$ with a spreading exponent $\alpha \approx 0.75$. This value is consistent with previous results for viscosity-limited spreading over a deep bath. 
A single power law of rescaled distance as a function of rescaled time reasonably captures our experiments, as well as different geometries, miscibilities, and surface tension modifiers (solvents and surfactants). This result enables engineering the spreading dynamics of a wide range of liquid-liquid systems.
\end{abstract}


\section{Introduction}\label{sec:intro}
\begin{figure}
\begin{centering}
\subfloat[\label{fig:1A}]{}
\subfloat[\label{fig:1B}]{}
\subfloat[\label{fig:1C}]{}
\includegraphics[width=1.2\linewidth]{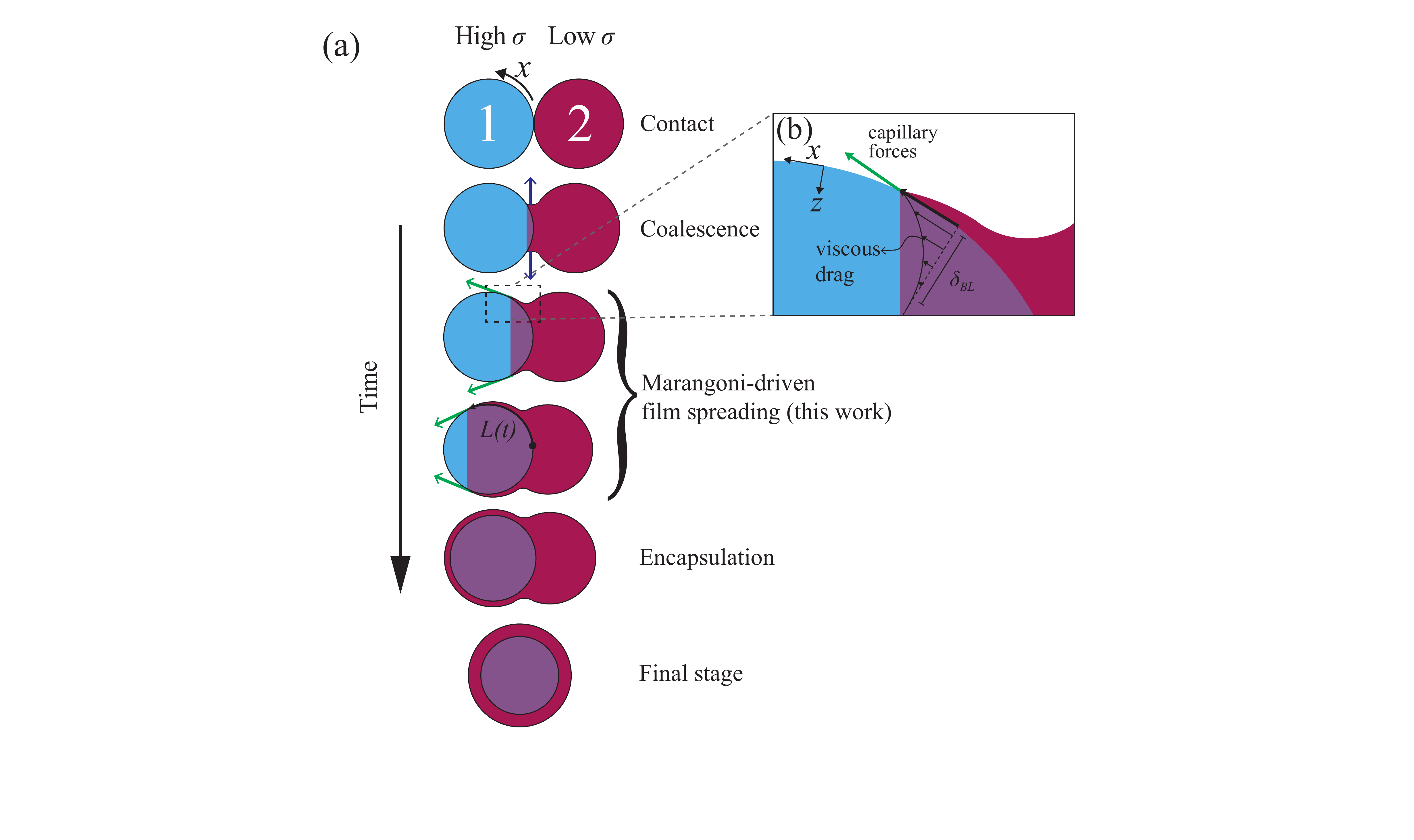}
\par\end{centering}
\caption{\label{fig:1}(Color online) Overview of binary drop spreading; $(a)$ At $t=0$ the drops touch. Initially, coalescence radially expands the neck due to local curvature as indicated by the blue arrows \citep{Paulsen2011}. This regime is followed by Marangoni-driven spreading of the drop with lower surface tension over the other one (green arrows). Ultimately, this mechanism results in encapsulation of drop 1. $(b)$ Indicative flows of Marangoni-driven spreading, where the Marangoni stress is balanced by a viscous boundary layer. 
}

\end{figure}

Liquids of low surface tension spread over liquid with high surface tension, which is known as Marangoni spreading. This phenomenon has been studied in various contexts, such as oil spills on the sea \citep{Franklin1774,Rayleigh1889,Fay1971,Hoult1972,Foda1980}, pulmonary surfactant replacement therapy \citep{Gaver1990,Matar2009}, and foam destruction \citep{Denkov2004}. Recently, Marangoni spreading has been used for encapsulation of a high-surface tension drop by a lower surface tension liquid, as shown in figure \ref{fig:1A}. This mechanism is used in pharmacy \citep{Yeo2003,Yeo2004}, for manufacturing of biomaterials \citep{Duan2013}, electronics \citep{TenCate2014}, food and vitamins \citep{Blanco-Pascual2014}, microparticles with multiple compartments \citep{Hayakawa2016,Kamperman2018}, and 3D printing of materials with controlled micro-architectures by in-air microfluidics \citep{Visser2018}.

The morphological outcome of two colliding drops in the air, such as encapsulation or breakup, has been assessed for miscible \citep{Yeo2003,Visser2018} and immiscible \citep{Chen2006a,Chen2007,Planchette2010,Focke2012} liquid pairs with different surface tensions. Encapsulation can also be achieved by impact of droplets with different sizes \citep{Liu2013} or different viscosities \citep{Focke2013}. However, to our best knowledge, their surface-tension-driven encapsulation dynamics have hardly been visualized. Drops can also be encapsulated by gentle deposition onto a bath with a lower surface tension, but here the film dynamics were only assessed during coalescence (an earlier regime that precedes encapsulation) \citep{Thoroddsen2007} or for surface tension induced necking \citep{Blanchette2009a,Sun2018}, rather than encapsulation. Encapsulation in the binary droplet geometry was studied for submerged droplet pairs, revealing a constant velocity of the spreading film both experimentally \citep{Nowak2016,Nowak2017} and numerically \citep{Blanchette2010,Liu2013}. However, it is unclear whether this result also applies to drop pairs in air, since the viscosity of the surrounding liquid plays an important role.

\captionsetup[subfigure]{labelformat=empty}

Knowledge of Marangoni spreading over a flat liquid surface with a higher surface tension could also provide clues to describe spreading over drops. This topic has been studied in many configurations, of which most can be classified according to four criteria: (i) pure liquid-driven versus surfactant-driven spreading, (ii) miscible versus immiscible liquid pairs, (iii) shallow versus deep liquid ``carrier'' layers, and (iv) spreading from a finite reservoir versus a source. Here, we focus on spreading of ethanol/water mixtures over pure water drops, corresponding to surfactant-free and miscible liquid pairs. For a droplet pair, the transition between deep and shallow carrier layers may depend on the thickness of the flow-induced viscous boundary layer as sketched in figure \ref{fig:1B}. Deep-layer behavior is expected if the boundary layer thickness $\delta_{BL}<D_1/4$, with $D_1\approx 2$~mm the inner drop's diameter \citep{Vernay2015}.
Using typical values for the density $\rho = 1000~$kg/m$^3$, viscosity $\eta= 1$~mPa s, and time $t=10$~ms, we obtain $\delta_{BL} = (\eta_{1}t/\rho_{1})^{1/2}\approx 0.1$~mm. Therefore, spreading over a deep layer is considered. Finally, the outer drop is assumed to be an infinite source, as its volume suffices to form a thick 
film around the inner drop. This configuration was first studied by Suciu et al. \citep{Suciu1967,Suciu1969,Suciu1970,Ruckenstein1970}, but for a quasi-steady regime that follows the initial expansion of the film that is relevant to drop encapsulation. 

The spreading distance $L(t)$ of a low-surface tension liquid over a liquid with a higher surface tension can be described by a power law \citep{Fay1971}: 
\begin{equation} \label{generalPowerLaw}
L(t)= a \beta t^{\alpha}.
\end{equation}
The spreading exponent $\alpha$, the prefactor $\beta$, and constant $a$ are usually reported as a function of the geometric and material parameters \citep{Jensen1995}, and are also the scope of this study. The canonical result for spreading on a deep bath is $\alpha = 3/4$ and $\beta = S^{1/2} (\rho \eta)^{-1/4}$, in which $S\approx \Delta \sigma$ represents the spreading parameter for liquid pairs in air, $\Delta \sigma = \sigma_1-\sigma_2$ is the surface tension difference between the liquids. These values follow from balancing the surface tension gradient with dissipation in the viscous boundary layer that develops while spreading on a deep layer \citep{Fay1969a,Hoult1972,Joos1977,Foda1980,Berg2009}, and were validated for immiscible, non-evaporative liquids \citep{Dussaud1998,Camp1987}, immiscible surfactant solutions \citep{Joos1985}, and spreading over insoluble surfactants \citep{Bergeron1996}. This scaling is maintained for immiscible microdroplets spreading over free-flowing thin films \citep{Vernay2015}. For miscible surfactant solutions, the spreading exponent is maintained around $\alpha = 0.75$ for low solubility \citep{Roche2014,Wang2015b} but it can drop to $\alpha = 0.4$ for highly soluble surfactants \citep{Tarasov2006}. An indicative value $a \approx 0.88$ applies to spreading of immiscible liquids in the radial geometry, but values in the range 0.665 to 1.52 have been reported \citep{Dussaud1998}.

For spreading of a pure low$-\sigma$ drop over a deep bath of a miscible liquid as considered here, a spreading exponent in the range $\alpha =0.53 \pm 0.03$ was measured for nitroethane, ethyl acetate \citep{Santiago-Rosanne2001}, and isopropanol drops \citep{Kim2017} deposited on water. Molecular dynamics simulations of ethanol solutions spreading over water revealed a similar exponent of $\alpha = 0.55 \pm 0.05$ \citep{Taherian2016}. These reduced values, as well as a decrease to $a \approx 0.3$, were attributed to dissolution of the spreading liquid into the bath by convective rolls that form at the film's edge \citep{Santiago-Rosanne2001,Kim2015}. Similarly, volatile films also reduce the spreading velocity, as evaporative cooling causes a density increase that results in vortex formation \citep{Dussaud1998}. An even lower exponent ($\alpha \approx 0.25$) was measured for ethanol drops on a water bath \citep{Dandekar2017}, and explained by balancing $\Delta \sigma$ with viscous dissipation within the spreading film as first reported by \cite{Bacri1996}. The applicability range of these earlier results is not yet clear. 

As the existing literature indicates that spreading exponents $1/4\lesssim\alpha\lesssim1$ could apply to the binary drop geometry, here we observe and quantify the Marangoni-driven spreading dynamics of miscible drop pairs. By encapsulating a fluorescent inner drop by an optically absorbing low-$\sigma$ liquid, we obtain the spreading distance as a function of time, the surface tension difference, and the viscosity. Subsequently, we determine the spreading exponents and compare these to systems with different geometries, surfactants and miscibilities. The paper is organized as follows: In section 2, the experimental setup an liquids are described. The results and discussion are described in section 3, followed by the conclusions in section 4.

\captionsetup[subfigure]{labelformat=empty}
\begin{figure}
\begin{centering}
\subfloat[\label{fig:2A}]{}
\subfloat[\label{fig:2B}]{}
\subfloat[\label{fig:2C}]{}
\subfloat[\label{fig:2D}]{}
\subfloat[\label{fig:2E}]{}
\par\end{centering}
\begin{centering}
\includegraphics[width=0.9\linewidth ,viewport={38bp 560bp 820bp 1150bp},clip]{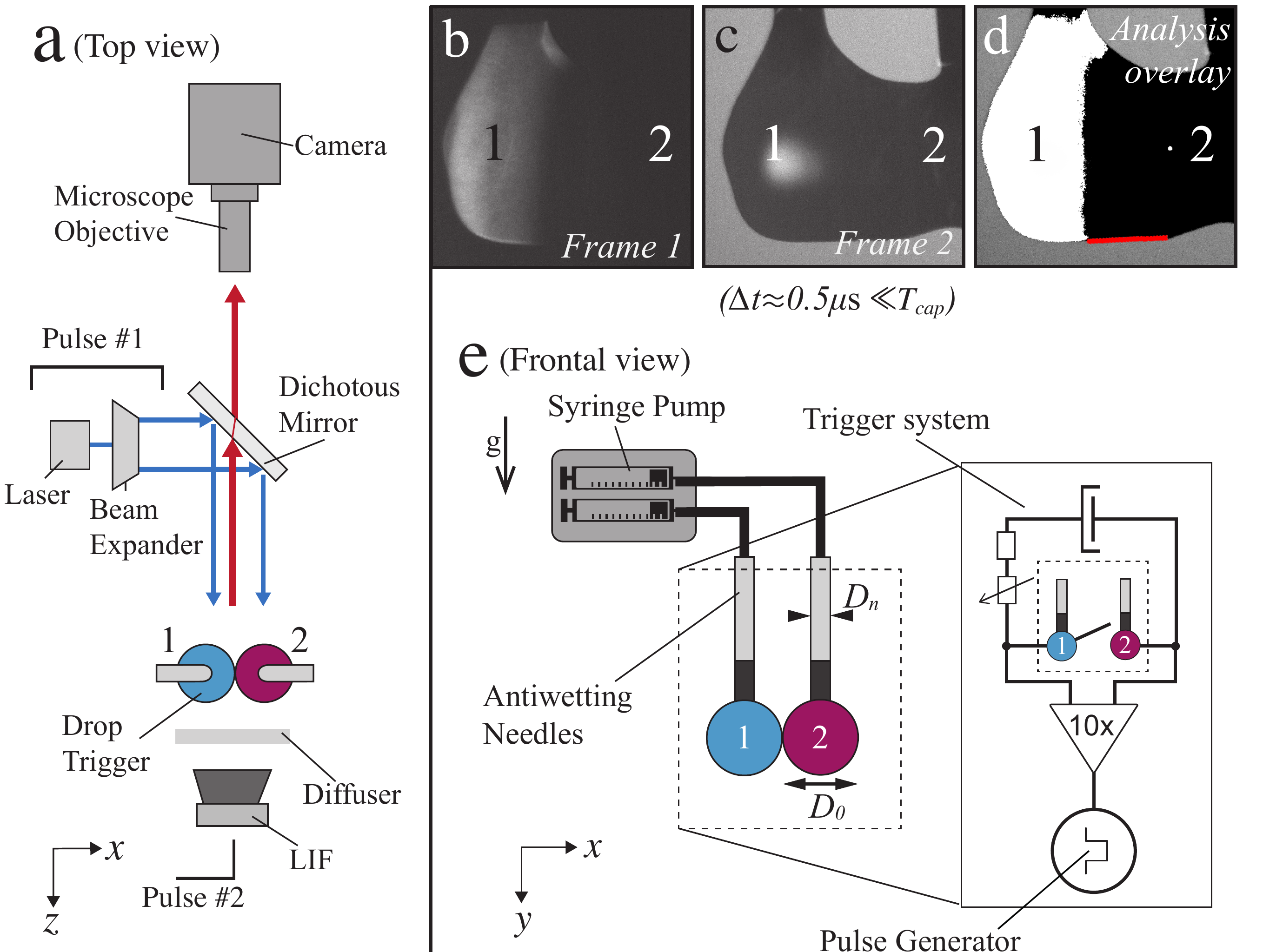}
\par\end{centering}
\caption{\label{fig:2}(Color online) $(a)$ Top view of the set-up. Drops 1 and 2 are illuminated by pulse \#1, as shown by the blue arrows. Fluorescent light is emitted only by drop 1 and passes the dichroic mirror, as indicated by the red arrow. $(b)$ Example fluorescent image. $(c)$ At virtually the same moment, pulse \#2 illuminates both droplets from the back. The resulting bright-field image is shown. $(d)$ Processed overlay of both images.  $(e)$ Setup from the side-view perspective of the camera. \textit{(Inset)} A pulse is generated when the drops closes an electric circuit. The delay time between this pulse and the image capture was controlled with a pulse generator.}
\end{figure}

\section{Experimental set-up and materials}\label{sec:setup}

\captionsetup[subfigure]{labelformat=empty}
\begin{figure}
\begin{centering}
\subfloat[\label{fig:3A}]{}
\subfloat[\label{fig:3B}]{}
\subfloat[\label{fig:3C}]{}
\subfloat[\label{fig:3D}]{}
\includegraphics[viewport=0bp 160bp 612bp 710bp,clip,width=0.9\linewidth]{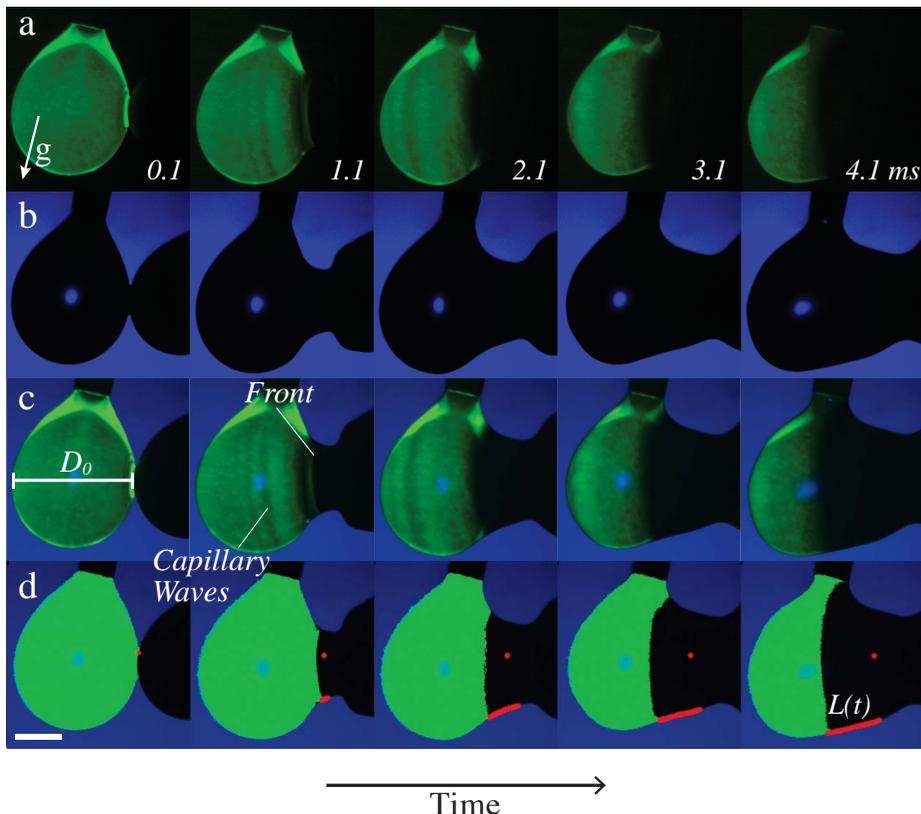}
\par\end{centering}
\vspace{-0.8cm}\caption{\label{fig:3}(Color online) Image analysis procedure. $(a)$ Typical darkfield image sequence, only showing fluorescent droplet 1. The numbers indicate the time after first contact in ms. $(b)$ Brightfield image sequence, showing the contours of both droplets.  $(c)$ Overlay of (a) and (b), revealing the spreading film. Capillary waves and spreading front are observed as indicated $(d)$ Result of the image analysis procedure. The red dot indicates the $x$-position of merging, while the red line indicates the spreading distance of the low-surface tension solution. The scale-bar indicates \SI{1}{\milli\meter}.}
\end{figure}

To create the binary drop geometry, two drops were suspended from teflon needles (Hamilton Company) that were fed by two identical syringes mounted on a syringe pump (Harvard PHD 2000). The needles were placed at a 2.5mm center-to-center distance, resulting in drops with a diameter $D_0 = 2.5\pm0.5$\si{\milli\meter}. 
The high-$\sigma$ drop consisted of milli-Q water as a base, to which fluorescein (emission at \SI{525}{\nano\meter}) was added for fluorescent visualization. The low-$\sigma$ drop consists of a $15\,\mathrm{vol}\%$ inkjet printer ink solution (Brother LC-800), to provide an optically absorbing film that blocks the fluorescent light of the high-$\sigma$ drop during spreading. The surface tension gradient was modified by adding ethanol to the low-$\sigma$ drop, resulting in a viscosity of $1.5 \pm 0.5$ \si{\milli\pascal\second}.   
The viscosity was adjusted by adding glycerol to one or both of the liquids. $1\,\mathrm{vol}\%$ NaCl solution was added to both liquids to increase the conductivity without significantly affecting the surface tension \citep{Schmid1962}. The material properties were obtained from the literature or measured (see supplementary materials). 

The visualization setup is depicted in figure \ref{fig:2A}. Stroboscopic imaging was used to generate two images of the drop pair at a controlled time after contact. The first image was illuminated with a pulsed laser (Litron Nano S PIV \SI{400}{\milli\joule}, wavelength \SI{532}{\nano\meter}, pulse duration 8ns), of which the optical path is shown by the blue arrows in figure \ref{fig:2A}. Only the fluorescent light is observed, as shown in figure \ref{fig:2B}. The second frame was exposed by diffuse illumination from behind both droplets, resulting in images of the droplet contours as shown in figure \ref{fig:2C}. Here, the pulse was provided by a second pulsed laser (Evergreen \SI{600}{\milli\joule}, wavelength \SI{532}{\nano\meter}) that was diffused with a fluorescent diffuser (LaVision) to prevent fringes. The delay time between both laser pulses was set to 500~ns, and the corresponding images were captured in separate frames of a dedicated dual-frame camera (PCO Sensicam qe). As this delay time is approximately 4 orders of magnitude shorter than the capillary time scale ($T_{cap}=\sqrt{\rho R_0^3/\sigma}\approx\SI{5}{\milli\second}$), no significant motion occurs between frames 1 and 2. The frames were overlaid (with excellent spatial collapse) as shown in figure \ref{fig:2D}, revealing the spreading extent of the film and the outer contour of the drops. Time series were generated by repeating the above procedure for different delays between the moment of drop-drop contact ($t=0$) and image capture. The moment of contact was obtained by closing an electrical circuit with the conductive drops, as shown in Figure \ref{fig:2E} (inset).

Figures \ref{fig:3A} and \ref{fig:3B} show example time series of the fluorescent drop and both drops' contours, respectively. The overlay in figure \ref{fig:3C} reveals the spreading extent of the film. An intensity threshold of $<5\%$ as compared to the uncovered (green) drop was chosen to determine the covered part with automated image analysis, corresponding to a film thickness of 63 \si{\micro\meter}. The exact value of this threshold had a minor influence on the spreading distance (see SI figure 6). Still, we would like to stress that we measure the spreading of relatively thick films that are relevant to encapsulation, rather than micrometer- or nanometer-thin films as reported previously \citep{Ruckenstein1970,Mann1991}. We traced the spreading along the bottom of the drop pair to prevent errors due to out-of-plane motion, as indicated by the red lines in figures \ref{fig:2D} and \ref{fig:3D}. To reduce the risk of errors, we averaged three measurements of the spreading distance for each configuration and each time step, and performed scans of the control parameters $\Delta\sigma$ and $\eta$ over the largest feasible range for which spreading still occurs.

\begin{figure}
\begin{centering}
\includegraphics[width=1\linewidth]{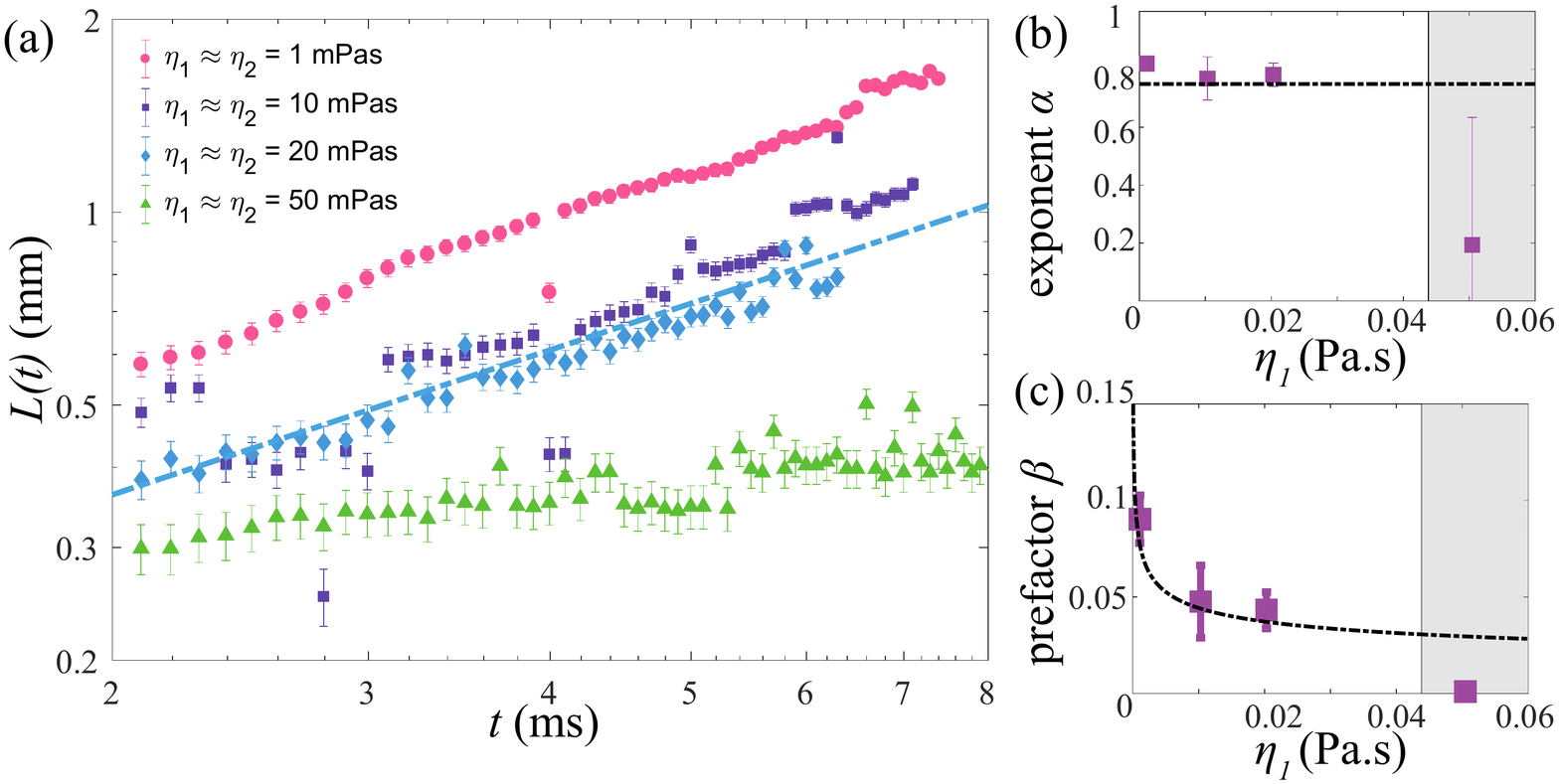}
\par\end{centering}\vspace{-2.5cm}
\caption{\label{fig:5}(Color online) (a) Time evolution of the leading edge position $L(t)$ as a function of the viscosity, with $18.9\leq\Delta\sigma\leq23.2$\si{\milli\newton\per\meter}. The dash-dotted line indicates $L(t) = 0.6 \beta t ^\alpha$ with $\alpha = 3/4$ and $\beta = \Delta\sigma^{1/2} (\rho_1 \eta_1)^{-1/4} $, which are expected for spreading over a deep bath. (b) The spreading exponent $\alpha$ as a function of the viscosity; the dash-dotted line indicates $ 3/4$. (c) The prefactor $\beta$ as a function of the viscosity, with the dashed line indicating $0.6\Delta\sigma^{1/2} (\rho_1 \eta_1)^{-1/4} $. The shaded areas indicate $Oh>0.2$.}
\end{figure}

\section{Results and discussion}\label{sec:results}

The position of the spreading front $L(t)$ was measured as a function of time and viscosity, as shown in figure \ref{fig:5}, revealing power-law behavior with an approximate scaling exponent $\alpha = 3/4$ for low viscosities ($\eta_1 \approx \eta_2 \leq  20$). The prefactor is reasonably described by $a=0.6$, i.e. $L(t) = 0.6\Delta\sigma^{1/2}(\rho\mu)^{1/4}t^{3/4}$. Increasing the viscosity to $\eta_1\approx\eta_2=50$\si{\milli\pascal\second} leads to a significant decrease in the spreading rate, and increasing the viscosity even further prevents the spreading. The Ohnesorge number for this case is $Oh = \eta/\sqrt{\rho \Delta\sigma D} \approx 0.2$, \textit{i.e.} spreading seems to be inhibited when global viscous forces become comparable to surface tension forces. 

\captionsetup[subfigure]{labelformat=empty}
\begin{figure}
\begin{centering}
\subfloat[\label{fig:4A}]{}
\subfloat[\label{fig:4B}]{}
\includegraphics[width=1\linewidth]{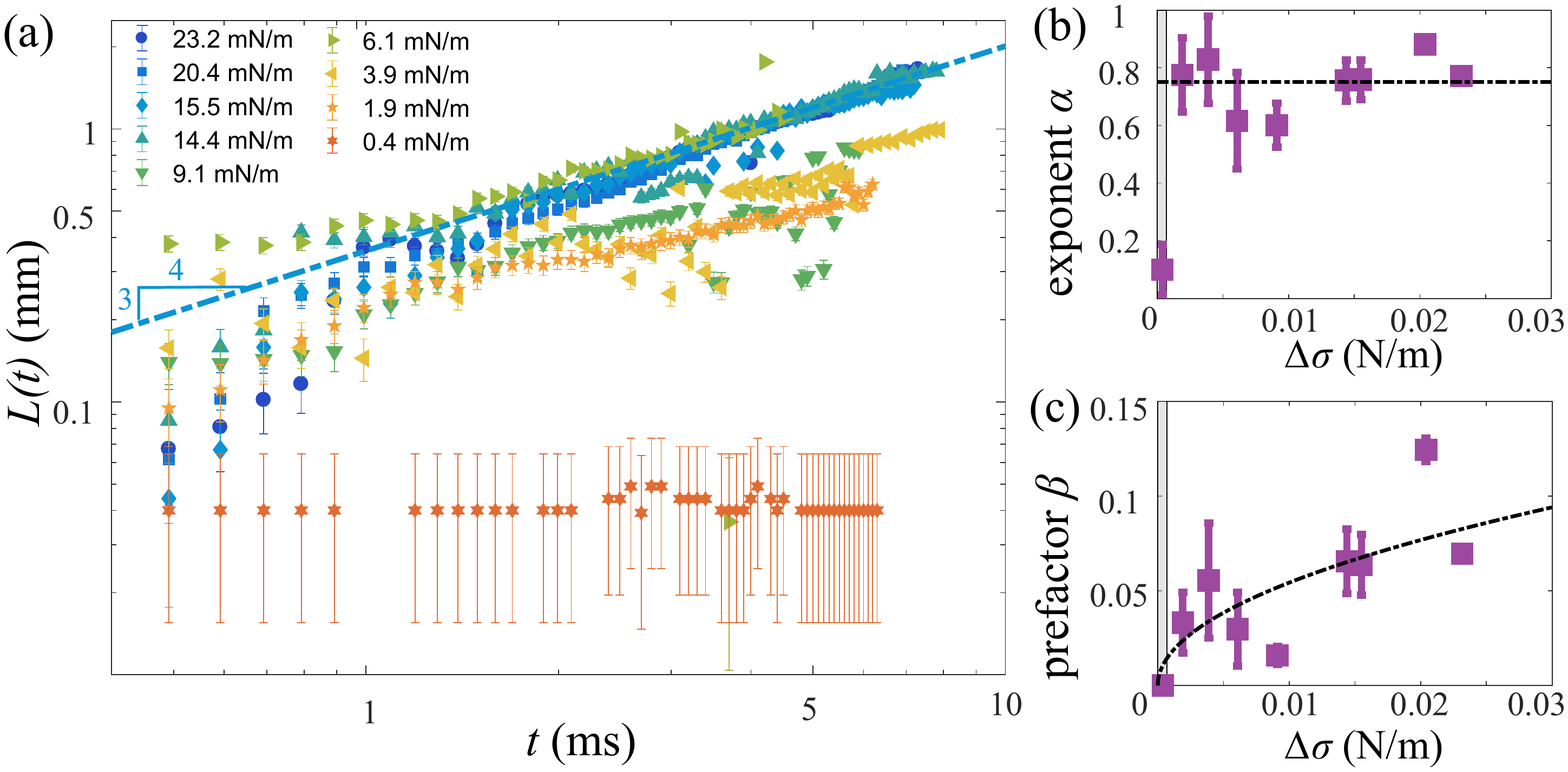}
\vspace{-3cm}
\par\end{centering}
\caption{\label{fig:4}(Color online) (a) Time evolution of $L(t)$ as a function of $\Delta\sigma$. 
(b) The exponent $\alpha$ as a function of $\Delta \sigma$. (c) Prefactor $\beta$ as a function of $\Delta \sigma$. In (b,c), the shaded areas close to the origin indicate $Oh>0.2$. The lines correspond to the equations in figure \ref{fig:5}.}
\end{figure}

Figure \ref{fig:4A} shows the spreading distance as a function of the surface tension difference, which was varied from $\Delta \sigma=0.4 $\si{\milli\newton\per\meter} to $\Delta \sigma=23.2$ \si{\milli\newton\per\meter}. 
The spreading exponents are still consistent with $\alpha = 3/4$, as shown in figure \ref{fig:4B}. The prefactors exhibit substantial statistical errors and data scattering, but a value of $0.6\Delta\sigma^{1/2}(\rho\mu)^{1/4}$ still reasonably captures them as shown in figure \ref{fig:4}c. Spreading is inhibited for $\Delta\sigma = 0.4 $\si{\milli\newton\per\meter}, for which $Oh \approx 0.15$. 

\begin{figure}
\begin{centering}
\subfloat[\label{fig:6A}]{}
\subfloat[\label{fig:6B}]{}
\includegraphics[width=1\linewidth]{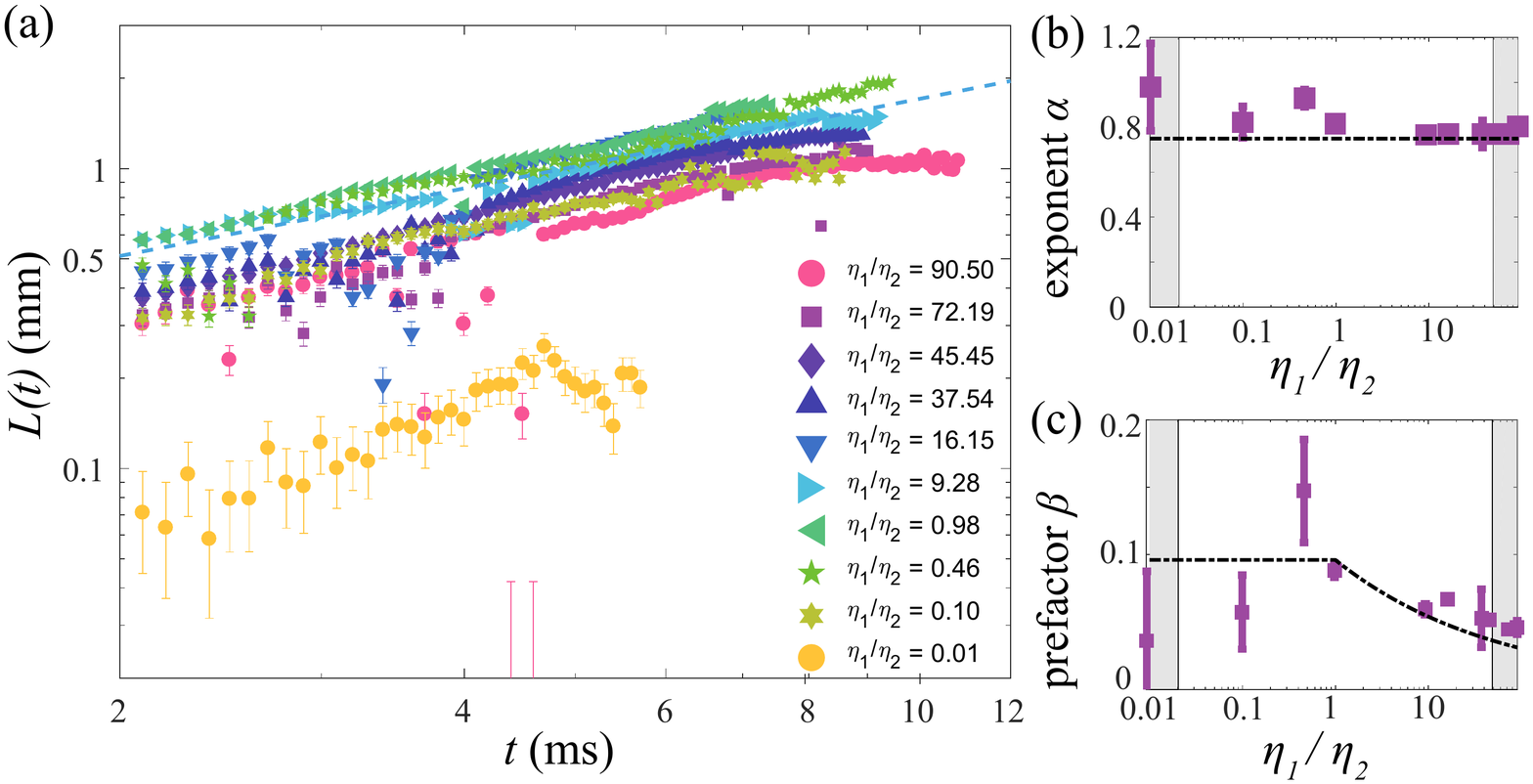}
\vspace{-3cm}
\par\end{centering}
\caption{\label{fig:6}(Color online) (a) Time evolution of the spreading edge $L(t)$ for various viscosity ratios $\eta_{1}/\eta_{2}$, with $19.8\leq\Delta\sigma\leq28.3$\si{\milli\newton\per\meter}. (b) Spreading exponent $\alpha$ and (c) prefactor $\beta$ as a function of $\eta_{1}/\eta_{2}$. The lines correspond to the equations in figure \ref{fig:5}.}
\end{figure}

The effect of changing the viscosity ratio $\eta_1/\eta_2$ between the drops is shown in figure \ref{fig:6A}. The fastest spreading is observed for a ratio of unity ($\eta_1=\eta_2\approx 1.5~$mPa s), as the viscosity of both liquids is set to their lowest values for this case. Increasing the viscosity of either liquid results in a decrease in spreading over the entire temporal domain. The spreading exponent (figure \ref{fig:6}b) is constant around $\alpha = 3/4$. The reduction in spreading is captured by the prefactor, which is consistent with theory for viscosity ratios $\eta_1/\eta_2\geq1$ (figure \ref{fig:6}c). The prefactor could not be reliably measured for low viscosity ratios $0.1<\eta_1/\eta_2<1$, but figure \ref{fig:6}a shows that the observed results lie only slightly below $\eta_1/\eta_2$. For even lower viscosity ratios ($\eta_1/\eta_2=0.01$), spreading is strongly suppressed (see figure \ref{fig:6}a). Here, the Ohnesorge number based on the more viscous liquid becomes of order 1 ($Oh \approx 0.5$).

\captionsetup[subfigure]{labelformat=empty}

\begin{figure}
\begin{centering}
\includegraphics[width=1\linewidth ,viewport=38bp 10bp 790bp 570bp,clip]{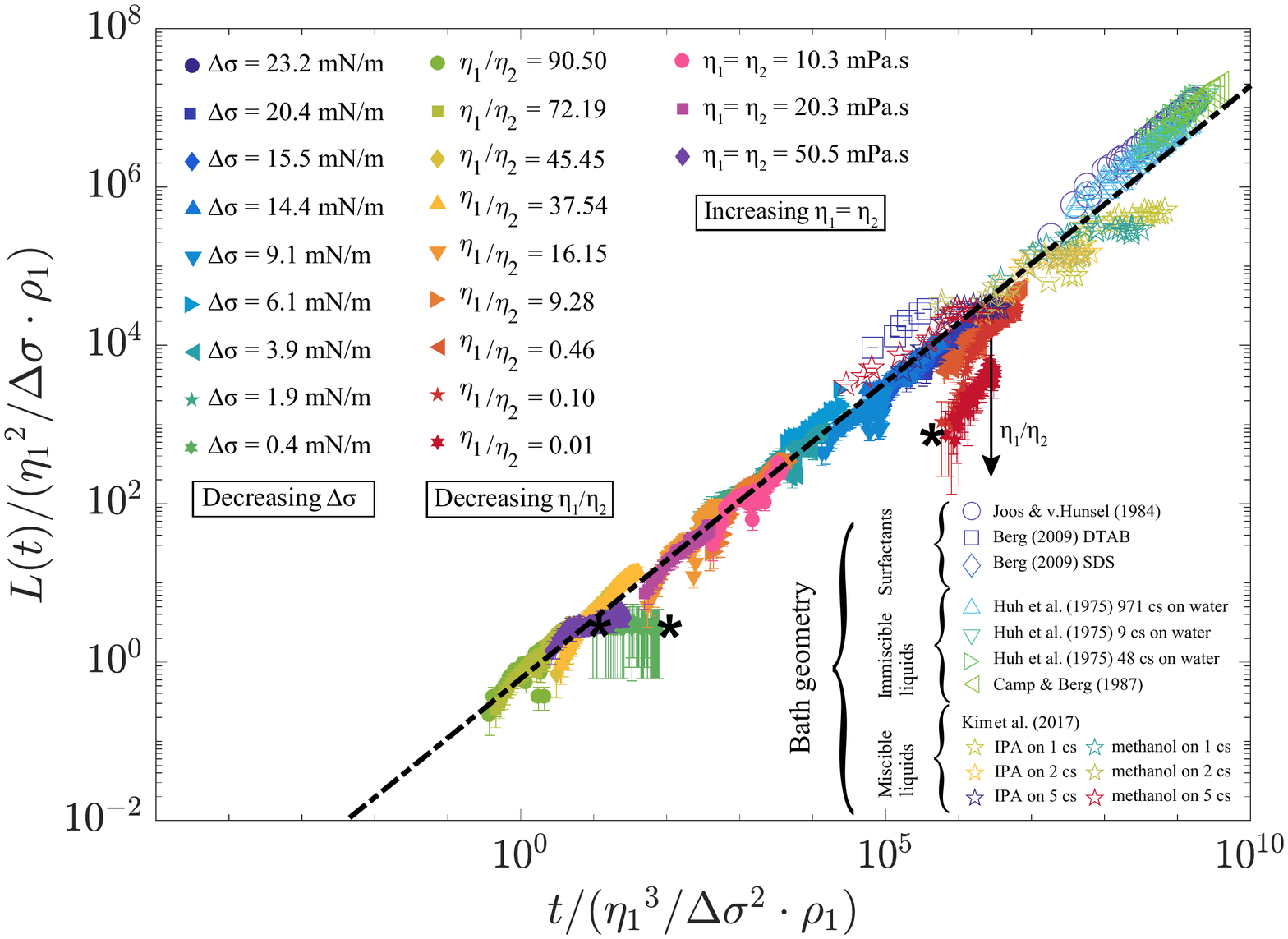}
\par\end{centering}
\caption{\label{fig:8}
(Color online) Rescaled spreading distance as a function of rescaled time, for our measurements and literature data for different geometries, miscibility, and surfactants. The dash-dotted line indicates  $\tilde{L} =0.6\tilde{t}^{~3/4}$, and the stars ({\bf *}) indicate measurements for which $Oh>0.15$.  
}

\end{figure}
As a final step, we rescale our results and compare these to Marangoni-driven spreading in other geometries, with different surfactants, and with immiscible liquids. The dimensionless spreading time and distance were formulated as proposed by \cite{Huh}:
\begin{eqnarray}
\tilde{t} & = & \frac{t} {\frac{\eta_{1}^{3}}{\Delta\sigma^{2}\rho_{1}}  },\\
\tilde{L} & = & \frac{L} {\frac{\eta_{1}^{2}}{\Delta\sigma\rho_{1}}  }.\\
\label{eq:dimlesT}
\end{eqnarray}
Our measurements are generally described by 
\begin{eqnarray}
\tilde{L} =0.6\tilde{t}^{~3/4},
\label{eq:mastercurve}
\end{eqnarray}
 as plotted in figure \ref{fig:8}. Measurements with $\eta_1/\eta_2<1$ do not collapse, as the viscosity of the outer drop is not included in the rescaled variables $\tilde{L}$ and $\tilde{t}$. (Attempts to accommodate this parameter did not result in their collapse to the master curve.) Figure \ref{fig:8} also shows that the same power law reasonably well describes drop spreading on a bath for surfactant-driven \citep{Joos1985, Berg2009}, immiscible \citep{Huh, Camp1987}, and miscible liquids \citep{Kim2017}. For unidirectional spreading, a higher constant in the range $0.66<a<2$, but typically $a\approx 1.39$, is expected \citep{Foda1980,Huh1975,Dussaud1998}. Therefore, the spreading distance in figure \ref{fig:8} is enhanced for the measurements by \cite{Joos1985, Berg2009,Huh} and \cite{Camp1987}. As discussed, constants $a\approx0.88$ and $a\approx0.3$ were reported for immiscible \citep{Dussaud1998} and miscible \citep{Santiago-Rosanne2001,Kim2015} liquids, respectively. The lower value was attributed to liquid densification by dissolution, which results in vortex formation by gravitational "sinking" of the heavier liquid \citep{Santiago-Rosanne2001}. Gravity will play a much smaller role in our experiments, due to their smaller size and spherical geometry, but vortices may still form as these were observed even in molecular dynamics simulations that do not include gravity \citep{Taherian2016}. Therefore, they are likely to develop in our system as well, possibly resulting in $a=0.6$. Even when ignoring these differences in constant $a$, a reasonable consistency exists between all measurements and equation \ref{eq:mastercurve}. 

Some of our results cannot be readily explained by this equation or the existing literature. For example, for low surface tension differences ($\Delta\sigma<4$~mN/m in figure \ref{fig:4}), a reduction in the spreading exponent is observed after a few milliseconds. This observation could indicate a transition to lower exponents in the range $0.25<\alpha<0.5$, as previously observed for miscible systems \citep{Santiago-Rosanne2001,Dussaud1998,Dandekar2017}. Vortex formation may play a role, as discussed above, but the role of dissipation in the film is also unclear. Scaling arguments that compare the viscosity in the film to that of the liquid substrate, i.e. the bath, predict a crossover from film- to substrate-limited spreading at a spreading radius $R_c = (\eta_2 V/\eta_1)^{1/3}$, with $V$ the droplet volume \citep{Bacri1996}. For most of our experiments, this criterion implies that spreading would be film-limited, for which $\alpha =1/4$. However, the generally observed spreading exponent for our measurements, $\alpha \approx 0.75$, suggests that dissipation in droplet 1, i.e. the substrate, is the dominant contribution. In contrast, film-limited spreading was recently reported for larger ethanol-water systems (remarkably, here $R>R_c$ so that substrate-limited spreading would be expected) \citep{Dandekar2017}. Although still limited in scope, these experimental results indicate that the transition between film-limited and substrate-limited spreading deserves attention. 

Furthermore, it is remarkable that measurements with reduced exponents are still reasonably captured by a power law with exponent 3/4. As yet, we have not found a physical explanation for this result, but the experiments by \cite{Kim2017} might prove a fruitful starting point for further investigation. As visible in figure \ref{fig:8}, their data for spreading over a bath with $\eta = 5$ mPa s are similar to our data, whereas lower viscosities deviate from equation (\ref{eq:mastercurve}). As such, these experiments may capture a transition between effectively immiscible and miscible spreading. 

Finally, we expected that the drop-drop geometry would play an important role, since the spreading film will meet itself at the back-side of the drop \citep{Tarasov2006} and a back-flow will develop within drop 1. This back-flow could result in meeting boundary layers that resemble spreading of thin films over a solid substrate, for which $\alpha =1/2$ \citep{Ahmad1972a,Hernandez-Sanchez2015}. However, the temporally sustained power-law behavior of most of our experiments suggests that these aspects play a minor role. In view of this geometry-invariance, investigating the above questions for short time scales in a flat geometry would directly contribute to the rational design of encapsulation by Marangoni-driven spreading.

\section{Conclusions}\label{sec:conclusion}
We experimentally studied the dynamics of Marangoni spreading of miscible liquids in the binary droplet geometry. Stroboscopic image sequences of mm-sized droplet pairs revealed the spreading distance as a function of time. 
The spreading distance is consistent with $L(t)\approx 0.6\Delta\sigma^{1/2} (\rho \eta)^{-1/4}  t^{3/4}$ for sufficient surface tension differences ($\Delta \sigma \gtrsim \SI{2}{\milli\newton\per\meter}$), low viscosities ($\eta_1\approx \eta_2<50$ mPa s), and moderate viscosity ratios ($0.1\lesssim\eta_1/\eta_2\lesssim 10$). Marangoni-driven encapsulation is suppressed when viscous forces become comparable to capillary effects, at an experimentally determined threshold of $Oh \gtrsim 0.2$. Non-dimensionalizing our results and literature experiments for different geometries, surfactants, and miscibilities revealed that a single power law reasonably captures all data, as shown in figure \ref{fig:8}. This curve may be therefore exploited to estimate the encapsulation time scale in various encapsulation applications. Several deviations were revealed as well, and discussed in view of existing data as well as further research required to predict the spreading dynamics of miscible liquids.

\section*{Acknowledgments}
We gratefully acknowledge insightful discussions with S. Wildeman, S. Karpitschka, J.H. Snoeijer and T. Kamperman, and we thank E. Zijlma for his contributions to the dedicated electronics.

\bibliographystyle{jfm}

\bibliography{library}

\end{document}